\documentclass[journal]{IEEEtran}
\usepackage{fullpage}
\usepackage{citesort}
\usepackage{epsfig}
\usepackage{graphicx}
\usepackage{caption}
\usepackage{subcaption}
\usepackage{fancybox}
\usepackage{color}
\usepackage{amssymb}
\usepackage{multirow}
\usepackage{hhline}

\renewcommand{\baselinestretch}{1.00}
\usepackage{amsthm}
\newenvironment{proofoutline}
 {\proof[Proof outline]}
 {\endproof }
 
\usepackage[cmex10]{amsmath}
\usepackage{amsfonts}
\usepackage{cases}
\usepackage{url}
\usepackage{epstopdf}

\newtheorem{theorem}{Theorem}
\newtheorem{proposition}{Proposition}

\newcommand{\cl}[1]{{\cal #1}}

\begin{document}
\title{A Universal Parallel Two-Pass
MDL Context~Tree Compression Algorithm}
\author{Nikhil~Krishnan,~\IEEEmembership{Student Member,~IEEE,}
        and~Dror~Baron,~\IEEEmembership{Senior~Member,~IEEE}
\thanks{N. Krishnan and D. Baron are with the Department
of Electrical and Computer Engineering, North Carolina State University, Raleigh,
NC, 27695 USA. E-mail:\{nkrishn,barondror\}@ncsu.edu.}
}
\maketitle
\begin{abstract}
Computing problems that handle large amounts of data necessitate 
the use of lossless data compression for efficient storage and transmission.
We present a novel lossless universal data compression algorithm
that uses parallel computational units 
to increase the throughput. The length-$N$
input sequence is partitioned into $B$ blocks.
Processing each block independently of the other blocks
can accelerate the computation by a factor of $B$,
but degrades the compression quality.
Instead, our approach is to first estimate the
minimum description length (MDL)
context tree source
underlying the entire input,
and then encode each of the $B$ blocks in parallel
based on the MDL source.
With this two-pass approach, the compression loss
incurred by using more parallel
units is insignificant.
Our algorithm is work-efficient, i.e.,
its computational complexity is $O(N/B)$.
Its redundancy is approximately
$B\log(N/B)$ bits above Rissanen's lower bound
on universal compression performance, with respect to any
context tree source
whose maximal depth is at most $\log(N/B)$.
We improve the compression by using different quantizers for states of the context
tree based on the number of symbols corresponding to those states.
Numerical results from a prototype implementation suggest that our algorithm 
offers a better trade-off between compression and throughput than competing 
universal data compression algorithms.
\end{abstract}


\IEEEpeerreviewmaketitle

\let\thefootnote\relax\footnote{This work was supported in part by the National Science Foundation under Grant CCF-1217749 and in part by the U.S. Army Research Office under Grants W911NF-04-D-0003 and W911NF-14-1-0314. Portions of this paper were presented at the IEEE International Symposium on Information Theory, 
Honolulu, Hawaii, June 2014~\cite{KrisBaron2014}, and at the IEEE Global Conference on Signal and Information 
Processing, Atlanta, Georgia, December 2014~\cite{KrisBaronSIP2014}.}

\section{Introduction} \label{sec:intro}

\subsection{Motivation} \label{subsec:intro:motivation}

The emergence of distributed cloud computing and big data problems raises new challenges in data storage and communication. In such distributed computing settings, the data may be processed remotely in clusters and the results are streamed to the end user through a network. The use of data streaming in big data problems makes it imperative to use fast lossless data compression algorithms whose primary features include good compression quality and high throughput.

Some applications of fast compression include internet backbone data compression and compression in high volume data generation applications such as scientific computing. 
Data can be compressed rapidly near the source of data generation, and can be transmitted through band limited channels. 
This compression scheme can reduce energy consumption and bandwidth requirements of the network. 

Several techniques are available to improve the throughput,
such as hardware acceleration~\cite{ArmingIBWT}, algorithmic approximations, and computer architecture optimizations~\cite{Lenhardt2012,Snappy,QuickLZ,FastLZ}. 
Although these acceleration, approximation, and optimization techniques may accelerate compression, 
there are many systems where these do
not suffice either due to limited speed up or poor compression quality.
Ultimately, in order for lossless compression
to become appealing for a broader range of applications,
we must concentrate
more on efficient new algorithms.

Over the last decade, inexpensive multi core processors such as 
{\em graphics processing units} (GPUs) have become available, 
and parallelization is a possible 
direction for fast compression algorithms. 
By compressing in parallel, we may obtain
algorithms that are faster by orders of magnitude.
However, with a naive parallel algorithm, which
consists of partitioning the original input into $B$ blocks and
processing each block independently of the other blocks,
increasing $B$ degrades the compression quality~\cite{FRT96}. Therefore,
naive parallel compression has limited potential.
Sharing information across blocks can improve the compression quality of data~\cite{Beirami2012}.

\subsection{Related work}

Stassen and Tjalkens~\cite{ST2001} proposed a
parallel compression algorithm based on context tree
weighting~\cite{Willems1995CTW} (CTW),
where a common finite state machine (FSM)
determines for each symbol
which processor should process it. Since the FSM processes the
original length-$N$ input in $O(N)$ time, Stassen and Tjalkens' method
does not support scalable data rates.

Franaszek et al.~\cite{FRT96} proposed a
parallel compression algorithm, which is related to LZ77~\cite{Cover06}, 
where the construction of a dictionary is
divided between multiple processors.
Unfortunately, the redundancy
(excess coding length above the entropy rate) of LZ77 is high.

Finally, Willems~\cite{Willems2000} proposed
a variant of CTW with $O(ND/B)$ time complexity, where $D$
is the maximal context depth that is processed.
Unfortunately, Willems' approach will not compress as well as CTW,
because probability estimates will be based on
partial information in between synchronizations
of the context trees.

\subsection{Contributions}

This paper presents a novel minimum description 
length~\cite{Rissanen1978} 
(MDL) data compression algorithm that coordinates
multiple computational units running in parallel,
such that the compression loss incurred by
using more computational units is insignificant.
Our main contributions are 
({\em i}) our algorithm is {\em work-efficient}~\cite{CLR}, 
i.e., it compresses $B$ length-($N/B$) blocks in
parallel with $O(N/B)$ time complexity;
({\em ii}) the redundancy of our algorithm is approximately $B\log(N/B)$ bits above the lower bounds on the best
achievable redundancy;
({\em iii}) 
we improve the compression quality by using different quantizers for states of the context
tree based on the number of symbols corresponding to those states;
({\em iv}) using a serial (non-parallel)
prototype implementation, we compare the compression and throughput as a function of the number of parallel computational units available; and
({\em v}) our algorithm has the useful property of random access~\cite{Kreft2010}, where any part of the compressed file can be decompressed without decompressing the entire file.
Numerical results show that our {\em parallel two-pass MDL} 
(PTP-MDL) algorithm provides a better trade-off between compression and throughput, which makes this algorithm attractive for big data problems.

The remainder of the paper is organized as follows. We review preliminary material in 
Section~\ref{sec:prelim}, describe our PTP-MDL algorithm~\cite{KrisBaron2014,BaronThesis} in Section~\ref{sec:basic}, discuss numerical
results in Section~\ref{sec:num_res}, and conclude in Section~\ref{sec:conclusion}.

\section{Data Compression Preliminaries} \label{sec:prelim}

\subsection{Universal data compression} 
\label{subsubsec:universal}

Lower bounds on the redundancy serve as benchmarks for compression quality. 
Consider length-$N$  sequences $x^N$ generated by a stationary ergodic source over a finite alphabet $\cl{X}$, i.e., $x^N \in\cl{X}^N$.
For an individual sequence $x^N$, the {\em pointwise redundancy} with respect to (w.r.t.) a class $\cl{C}$ of source models is
\begin{equation*}
\rho(x^N)\triangleq{l(x^N)-N\widehat{H_x}},
\end{equation*}
where $l(x^N)$ is the length of a uniquely decodable code~\cite{Cover06} for $x^N,$
and  $\widehat{H_x}$ is the entropy rate of $x^N$ w.r.t. the best model in $\cl{C}$
with parameters set to their {\em maximum likelihood}  (ML) estimates. Weinberger et al.~\cite{WMF1994}, whose result was improved upon later, proved for a source with $K$ (unknown) parameters that
\begin{equation}
\label{eq:WMF94}
\rho(x^N)  \geq \frac{K}{2}(1-\epsilon)\log(N),
\end{equation}
for any $\epsilon>0$,
except for a set of inputs whose probability vanishes
as $N\rightarrow\infty$, where $\log(\cdot )$ denotes the base-2 logarithm.
Similarly, Rissanen~\cite{Rissanen1996} proved that, for universal
compression of 
independent and identically distributed (i.i.d.) sequences, the worst case redundancy (WCR) is
\begin{equation}
\label{eq:RIS_IID}
\rho(x^N)  \geq \frac{|\cl{X}|-1}{2}\log(N)+C_{|\cl{X}|}+o(1),
\end{equation}
where $|\cl{X}|$ denotes cardinality of $\cl{X}$,  
the constant $C_{|\cl{X}|}$ was specified, 
and the $o(1)$ term vanishes as $N$ grows.
Rissanen's result for the WCR holds for any sequence whose ML estimates satisfy the central limit theorem, where we can replace $|\cl{X}|-1$ in the expression for the WCR (\ref{eq:RIS_IID}) of i.i.d. sequences with $K$.
Because i.i.d. models are too simplistic for modeling
real-world data, we use tree sources instead.

\subsection{Tree sources}

\renewcommand{\figurename}{Figure}

\begin{figure}[t]
\begin{center}
\includegraphics[angle=0,width=5.5 cm]{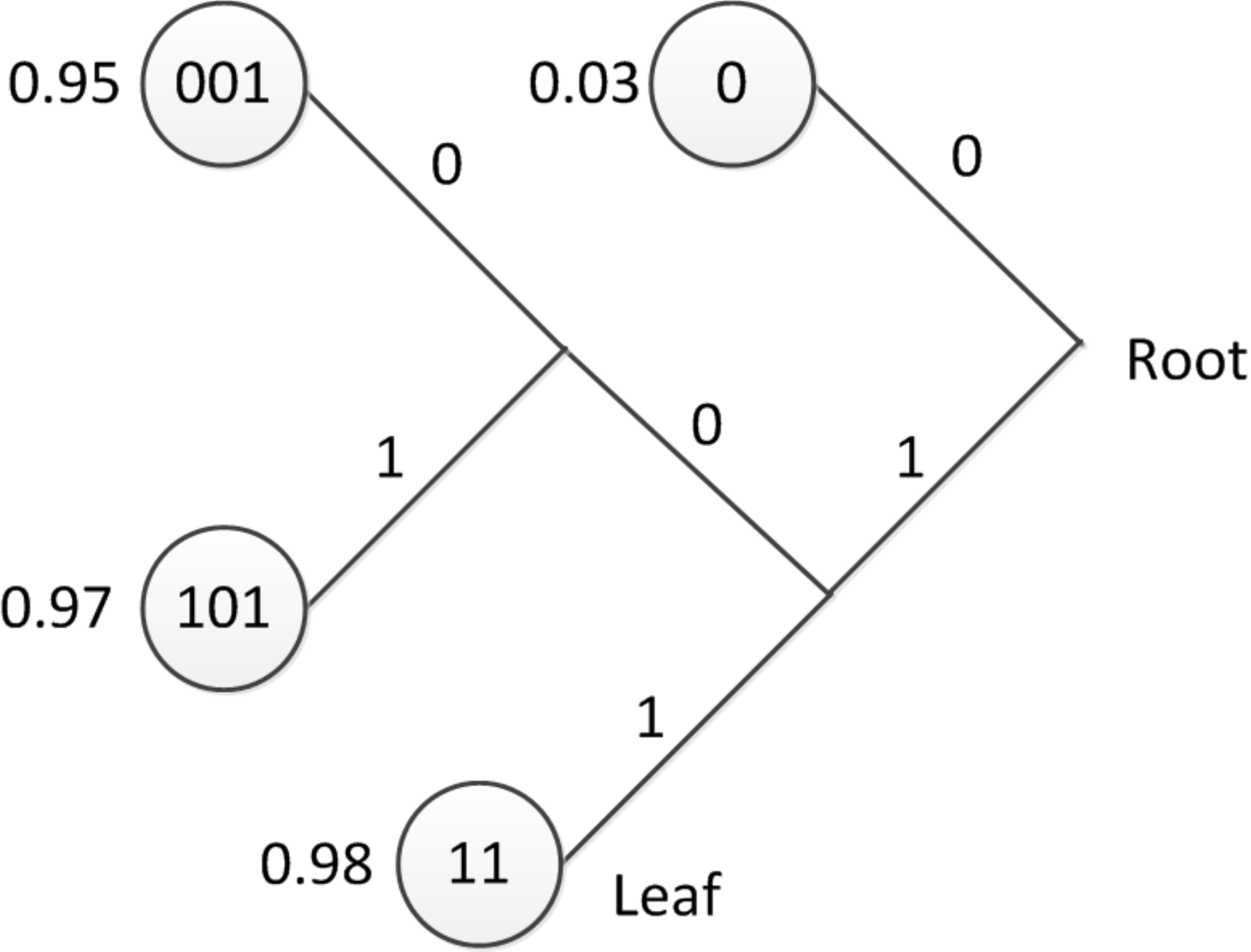}
\end{center}
\caption
{A tree source over $\cl{X}=\{0,1\}$.
The states are $\cl{S}=\{0,11, 001, 101\}$
and the conditional probabilities are $p(x_i=1|0)=0.03$,
$p(x_i=1|11)=0.98$, $p(x_i=1|001)=0.95$, and $p(x_i=1|101)=0.97$.
Data generated by this tree are quite compressible, because conditional probabilities are close to 0 and 1.}
\label{fig:tree}
\end{figure}

Let $x_i^j$ denote the {\em sequence}
$x_i,x_{i+1},\ldots,x_j$  where $x_k\in\cl{X}$
for $i\leq{k}\leq{j}$.
Let $\cl{X}^*$ denote the set of finite-length
sequences over $\cl{X}$.
Define a {\em context tree source}  $\{\cl{S},\Theta\}$~\cite{Willems1995CTW}
as a finite set of sequences called states
$\cl{S}\subset\cl{X}^*$ that is
complete and proper~\cite[p.654]{Willems1995CTW},
and a set of conditional probabilities $\Theta=\{p(\alpha|s):\ \alpha\in\cl{X},\ s\in\cl{S}\}$.
We say that $s$ {\em generates} symbols following it.
Because $\cl{S}$ is complete and proper, the sequences
of $\cl{S}$ can be arranged as leaves on
an $|\cl{X}|$-ary tree~\cite{CLR}
(Figure~\ref{fig:tree});  the
unique state $s$ that generated
$x_i$ can be determined by entering the tree at the root,
first choosing branch $x_{i-1}$, then branch $x_{i-2}$,
and so on, until some leaf $s$ is encountered.
Let $D\triangleq\max_{s\in\cl{S}}|s|$ be the {\em maximum context depth}.
Then the string
$x_{i-D}^{i-1}$ uniquely determines the current state $s$;
the previous symbols $x_{i-L}^{i-1}$ ($L\leq{D}$)
that uniquely determine the current state $s$ are called the
{\em context}, and $L$ is called the {\em context depth}
for state $s$.

\subsection{Semi-predictive and two-pass data compression}

{\bf Semi-predictive compression:}
Consider a tree source structure $\cl{S}$ whose explicit
description requires $l_{\cl{S}}$ bits,
and denote the probability of the input sequence $x^N$
conditioned on the tree source structure
$\cl{S}$ by $p_{\cl{S}}(x^N)$.
Using $\cl{S}$, the coding length required
for $x^N$ is 
$l_{\cl{S}} + l_{\cl{C}}$, where $l_{\cl{C}} = -\log(p_{\cl{S}}(x^N))$ is the coding length.
Let the MDL tree source structure $\cl{\widehat{S}}$
be the tree structure that provides the
shortest description of the data, i.e.,
\begin{equation*}
\cl{\widehat{S}} \triangleq
\arg\min_{\cl{S}\in\cl{C}}
\left\{l_{\cl{S}}+l_{\cl{C}}\right\},
\end{equation*}where $\cl{C}$ is the class of tree source models
being considered.
The semi-predictive approach~\cite{BWTMDL,Volf1995CTWMDL,WillemsCISS}
processes the input $x^N$
in two passes.
Pass~I first estimates $\cl{\widehat{S}}$
by context tree pruning, which minimizes the coding length.
The structure of $\widehat{\cl{S}}$ is then encoded explicitly.
Pass~II uses $\widehat{\cl{S}}$ to encode the
sequence $x^N$ sequentially, where the parameters $\widehat{\Theta}$ are estimated while encoding $x^N$.
The decoder first determines
$\widehat{\cl{S}}$, and afterwards uses it to decode $x^N$ sequentially from the two-pass code.

{\bf Two-pass compression:}
In contrast to the semi predictive approach, which only encodes the 
structure of $\cl{\widehat{S}}$ in Pass I, our two-pass
approach also encodes the parameter values $\widehat{\Theta}$.
We use a two-pass approach instead of
a semi-predictive approach,
because estimating $B$ sets of parameters in parallel,
one for encoding each of the $B$ blocks in Pass~II,
has $\rho(x^N) \approx 0.5B|\cl{S}|\log(N/B)$, 
whereas the two-pass approach has $\rho(x^N) \approx 0.5|\cl{S}|\log(N)$, 
and the latter redundancy is smaller. 
Despite the parallel nature of our algorithm, it incurs 
a single redundancy term for lack of knowledge of the parameters in Pass~I instead of $B$ redundancy terms in Pass~II.

\section{PTP-MDL Algorithm} \label{sec:basic}

In order to keep the presentation simple, we restrict
our attention to a binary alphabet, i.e., $\cl{X}=\{0,1\}$;
the generalization to non-binary alphabets is
straightforward.
We will show that PTP-MDL has $O(N/B)$ time
complexity when we restrict $D \leq \log (N/B)$, while still approaching the pointwise redundancy bound (\ref{eq:WMF94}).
These properties enable scalable data rates without
a factor-$B$ increase in the redundancy.

\subsection{Overview}

\begin{figure}[t]
\hspace*{-1mm}
\includegraphics[angle=0,width=8 cm]{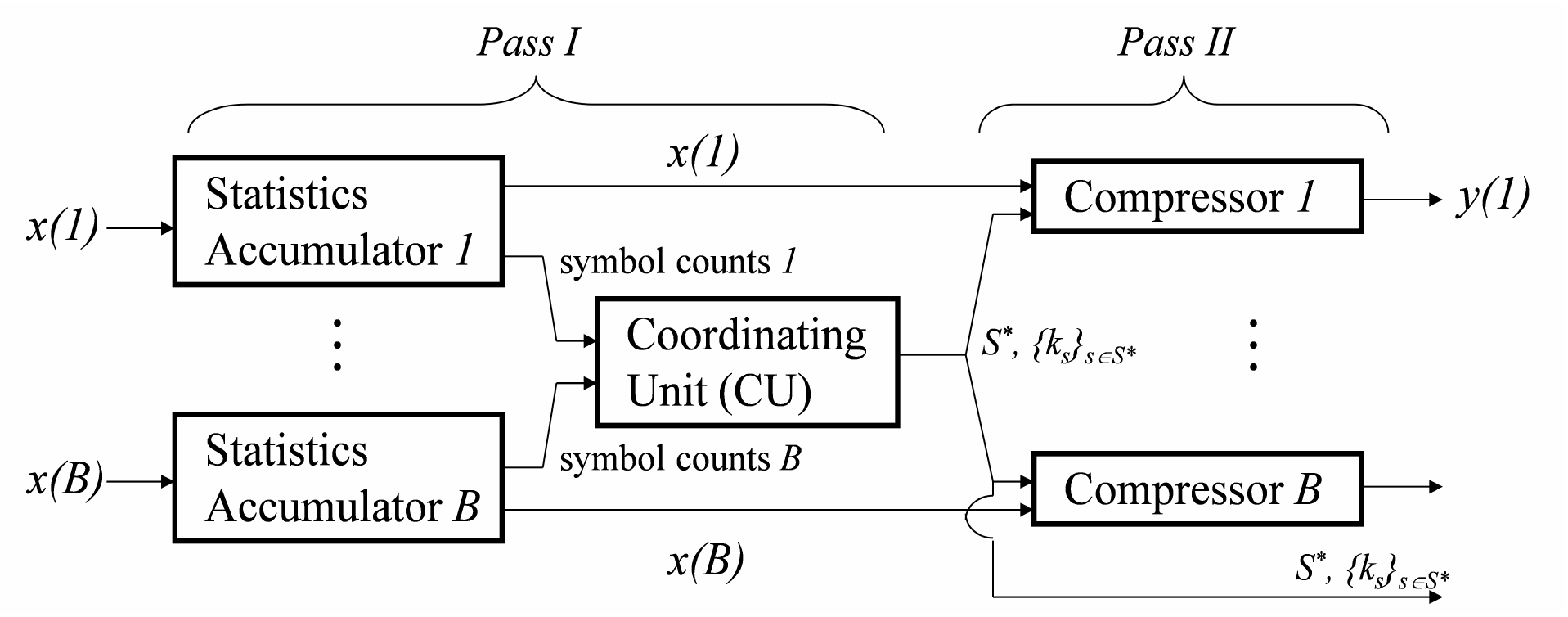}
\caption
{Block diagram of the PTP-MDL encoder.}
\label{fig:PTP-MDL}
\end{figure}

{\bf Encoder:}
A block diagram of the PTP-MDL encoder
is shown in Figure~\ref{fig:PTP-MDL}.
In Pass~I, the PTP-MDL encoder employs $B$ computational units called {\em parallel units} (PUs)
that work in parallel to accumulate statistical information
on $B$ blocks in $O(N/B)$ time, and a
{\em coordinating unit} (CU) that controls the PUs and
computes the MDL source estimate $\{\widehat{\cl{S}},\widehat{\Theta}\}$.

Without loss of generality, we assume $N/B \in {\mathbb Z}^+$.
Define the $B$ blocks as
$x(1)=x_1^{N/B},
x(2)=x_{N/B+1}^{2N/B},
\ldots,
x(B)=x_{N-N/B+1}^{N}$.
Parallel unit~$b$, where $b\in\{1,\ldots,B\}$,
first computes for each depth-$D$ context $s$
the {\em block symbol counts} $n^{\alpha}_s(b)$,
which are the number of times 
$\alpha$
is generated by $s$ in $x(b)$,
\begin{equation}
n^{\alpha}_s(b) \triangleq
\sum_{i=(b-1)(N/B)+ D+1}^{b(N/B)} 1_{\left\{x_{i-D}^{i}=s\alpha\right\}},
\alpha\in\cl{X},
\label{eq:def:n_s_alpha_b}
\end{equation}
where $s\alpha$ denotes concatenation of $s$ and $\alpha$,
and $1_{\{\cdot\}}$ is the indicator function.
For each state $s$ such that $|s|<D$, the CU
either retains the children states $0s$ and $1s$ in the
MDL source, or prunes them and only retains $s$,
whichever results in a shorter coding length.
Details of the pruning decision appear in Section \ref{subsec:basic:phaseI:MDL}.
Note that a single encoder compressing all $N$ symbols
has access to the last $D$ symbols from the previous 
block as context for encoding the first $D$ symbols of the current 
block (except for the first block). However, the MDL source
of a single encoder
is suboptimal in PTP-MDL, because this source does not reflect the actual symbols compressed by PTP-MDL.

In Pass~II, each of the $B$ blocks is compressed by a PU.
For each symbol $x_i(b)$, PU~$b$ first determines the 
generator state $G_i(b)$,
the state $s$ that generated the symbol $x_i(b)$.
PU~$b$ then assigns $x_i(b)$ a probability
according to the parameters that were
estimated by the CU in Pass~I, and sequentially feeds the
probability assignments to an arithmetic encoder~\cite{Cover06}.

\begin{figure}[t]
\includegraphics[angle=0,width=8 cm]{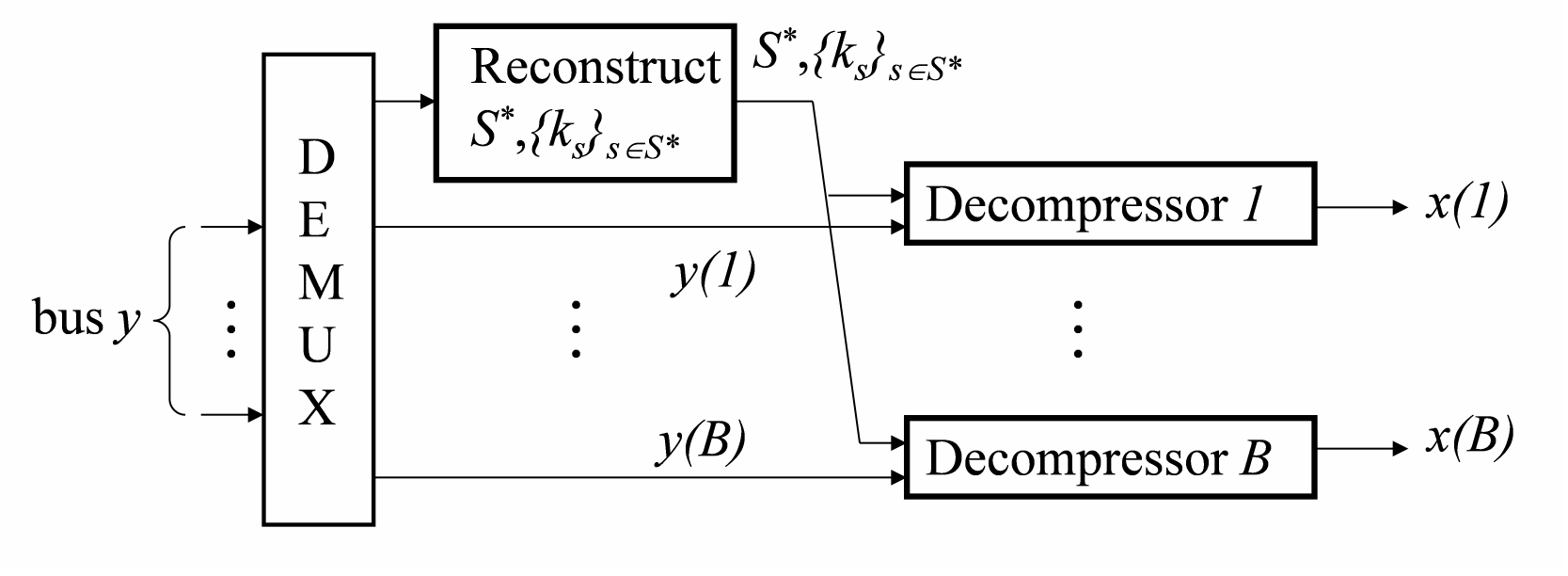}
\vspace{-1mm}
\caption
{Block diagram of the PTP-MDL decoder.}
\label{fig:PTP-MDLdec}
\end{figure}

{\bf Decoder:}
A block diagram of the PTP-MDL decoder
is shown in Figure~\ref{fig:PTP-MDLdec}.
The structure of the decoder is similar to that of Pass~II.
The approximated MDL source structure $\widehat{\cl{S}}$
and quantized parameters $\widehat{\Theta}$ are
first derived from the parallel source description
(see Section~\ref{subsec:basic:bitstream}).
Then, the $B$ blocks are decompressed by $B$ decoding blocks.
In decoding block~$b$, each symbol $x_i(b)$ is sequentially
decoded by determining the generator state $G_i(b)$, assigning
a probability to $x_i(b)$ based on the parameter
estimates, and applying an arithmetic decoder~\cite{Cover06}.

\subsection{Parallel source description} \label{subsec:basic:bitstream}

{\bf Two-pass codes in the PTP-MDL algorithm:}
Having received the block symbol counts $n^{\alpha}_s(b)$  from the PUs
(\ref{eq:def:n_s_alpha_b}), the CU computes the {\em symbol counts} 
generated by state $s$ in the entire sequence $x^N,$
\begin{equation}
n^{\alpha}_s = \sum_{b=1}^B n^{\alpha}_s(b),\qquad\alpha\in\cl{X}.
\label{eq:n^alpha_s}
\end{equation}
The CU can then compute the ML parameter estimates
of $p(1|s)$ and $p(0|s)$,
\begin{equation*}
\theta_s\triangleq \theta_s^1 = \frac{n_s^1}{n_s^0+n_s^1}
\quad \quad
\text{ and }
\quad \quad
\theta_s^0 = 1 - \theta_s^1,
\end{equation*}
respectively. 
The ML parameter estimates for each state $s$ are quantized into one of
\begin{eqnarray}
K_s  
&\triangleq&  
\left\lceil\sqrt{
2\pi^2\ln(2)\left(\frac{1}{2}-\frac{3}{16\ln(2)}\right)N}\right\rceil
\nonumber \\
&\approx&
\left\lceil 1.772\sqrt{N}\right\rceil
\label{eq:K^*}
\end{eqnarray}
{\em representation levels} based on Jeffreys' prior
such that each bin has the same probability mass~\cite{BaronBreslerMihcak2003},
where $\lceil \cdot \rceil$ denotes rounding up.
Jeffreys' prior for the scalar parameter $\theta$ is $p(\theta) \propto 1/{\sqrt{\theta (1-\theta)}}$, which is the arcsine distribution, and is almost flat for interior values of the parameter, $\theta \in (0,1)$. 
The representation levels and bin edges are computed using a closed form quantizer~\cite{BaronBreslerMihcak2003}.
The {\em bin index} and representation level for state $s$ are denoted by $k_s$ and $r_s$, respectively.
Denoting the quantized ML estimate of $\theta_s^{\alpha}$
by $\widehat{\theta}_s^{\alpha}$, we have $\widehat\theta_s^1=r_s$
and $\widehat\theta_s^0=1-r_s$. Recall that,
at the end of Pass~I, the CU
has computed the MDL structure estimate $\widehat{\cl{S}}$.
If $s\in\widehat{\cl{S}}$, then the first pass of the two-pass code
for symbols generated by $s$
consists of encoding $k_s$ with $\log(K_s)$ bits.
The WCR using this quantization approach is 1.047 bits per state above Rissanen's redundancy bound~\cite{BaronThesis,Rissanen1996,BaronBreslerMihcak2003}.

In Pass~II, which implements the second pass of
the two-pass code, each PU~$b$ encodes its
block $x(b)$ sequentially. For each symbol $x_i(b)$,
PU~$b$ determines the generator state
$G_i(b)$.
The symbol $x_i(b)$
is encoded according to the probability assignment
$\widehat{p}(x_i(b))\triangleq\widehat{\theta}_{G_i(b)}^{x_i(b)}$
with an arithmetic encoder~\cite{Cover06}.
Thus, the probability assigned by all $B$ PUs
to the symbols in $x^N$ whose generator state is $s$ is
\begin{equation}
\prod_{b=1}^B
\prod_{\left\{i:\ G_i(b)=s,\ i>D\right\}}
\widehat{p}(x_i(b)) =
(r_s)^{n_s^1}(1-r_s)^{n^0_s}.
\label{eq:length_part_two}
\end{equation}
Equation (\ref{eq:length_part_two})  provides the same
coding length for two-pass codes in a
parallel compression system as we would obtain in a serial
system~\cite{BaronThesis}.

{\bf Coding lengths in Passes~I and~II:}
In Pass~I, the structure $\widehat{\cl{S}}$ is described with
the {\em natural code}~\cite{Willems1995CTW}.
For a binary alphabet,
$|\text{natural}_{\widehat{\cl{S}}}|\leq{2}|\widehat{\cl{S}}|-1$ bits;
this is the {\em model redundancy} of PTP-MDL.
The parameters $\widehat{\Theta}$ are described as the $|\widehat{\cl{S}}|$ indices $k_s$ in the order in
which the leaves of $\widehat{\cl{S}}$ are reached in a depth-first
search~\cite{CLR}; this description can be implemented with
arithmetic coding~\cite{Cover06}.
The corresponding coding length
is the {\em parameter redundancy} of PTP-MDL.
We denote the length of the descriptions of $\widehat{\cl{S}}$
and $\widehat{\Theta}$ generated in Pass~I
by $l^I_{\cl{S}}$ bits. Using (\ref{eq:K^*}),
\begin{eqnarray}
l^I_{\cl{S}} &=&
|\text{natural}_{\cl{S}}|
+|\widehat{\cl{S}}|\log(K_s) \label{eqn:l1} \\
&\lessapprox&
\left[2|\widehat{\cl{S}}|-1\right]
+|\widehat{\cl{S}}|
\left[\log(1.772)+\frac{1}{2}\log(N)\right], \nonumber
\end{eqnarray}
where $\lessapprox$ denotes less than or approximately equal to.

In Pass~II, the coding length is mainly determined by symbol probabilities conditioned on generator states as given by (\ref{eq:length_part_two}).
There are two additional terms that affect the coding
length in Pass~II.
First, {\em coding redundancy} for each arithmetic encoder  with $\log(N)$
bits of precision requires $O(1) \leq 2$ bits~\cite{Cover06}.
Second, {\em symbols with unknown context}  at the beginning of $x(b)$; 
we encode the first $D$ symbols of each block $x(b)$ directly using $D$ bits per block.
Denoting the combined length of all $B$ codes
in Pass~II by $l^{II}_{\cl{C}}$ bits, we have
\begin{equation}
l^{II}_{\cl{C}}
\lessapprox
B\cdot{(D+2)}
-\sum_{s\in\cl{S}}
[n_s^1 \log(r_s) + n_s^0 \log(1-r_s)]
\label{eq:length:PhaseII}.
\end{equation}
Combining (\ref{eqn:l1}) and (\ref{eq:length:PhaseII}), we have the following result for the redundancy.
\begin{theorem}\label{theo1} (Theorem 18 in Baron~\cite{BaronThesis}.)
The pointwise redundancy of the PTP-MDL algorithm over the ML entropy of the input sequence $x^N$ w.r.t. the MDL source structure $\widehat{S}$, which is an element in the class of tree sources of depth $D \leq \log (N/B)$, satisfies
\begin{equation*}
\rho_P(x^N) \leq  B \left[ \log \left( \frac{N}{B}\right)  + 2  \right] + \frac{|\widehat{S}|}{2} \left[ \log \left(N \right) + O\left(1\right) \right],
\end{equation*}
where the subscript $P$ denotes the PTP-MDL compression algorithm.
\end{theorem}

\begin{proofoutline}
(For a detailed presentation, see pages 107 -- 115 in Baron~\cite{BaronThesis}.) The components of redundancy w.r.t. $\widehat{S}$ are as follows.

\begin{itemize}
\item The model redundancy of the natural code is at most $2|\widehat{S}| - 1$ bits. 
\item From Section~\ref{subsubsec:universal}, we saw that Rissanen's bound for the WCR (\ref{eq:RIS_IID}) is $\frac{1}{2} \log(N) + O(1)$ bits per state. The parameter redundancy introduced by two-pass codes based on Jeffreys' prior described in Section~\ref{subsec:basic:bitstream} is within $1.047$ bits per state above Rissanen's redundancy bound~\cite{BaronBreslerMihcak2003}.
\item Each PU $b$ encodes the first $D = \log \left(\frac{N}{B} \right) + O(1) $ bits directly.
\item Because the arithmetic computations are performed with finite precision, the outcome $\widehat{S}$ may not be an MDL tree source structure for $x^N$. Theorem~15 in Baron~\cite{BaronThesis} indicates that the coding length with $\widehat{S}$ is at most $O(1)$ bits more than the coding length obtained from the MDL context tree source structure for $x^N$.
\item The upper bound of arithmetic coding redundancy is 2 bits per PU~\cite{Cover06}.
\end{itemize}
In combination, these steps yield the WCR result of Theorem~\ref{theo1}. $\blacksquare$
\end{proofoutline}

The pointwise redundancy of naive parallel compression, which is denoted by $\rho_N(x^N)$ and is defined in Section~\ref{subsec:intro:motivation}, is worse compared to $\rho_P(x^N)$.

\begin{proposition}
The pointwise redundancy of the naive parallel compression algorithm over the ML entropy of the input sequence $x^N$  w.r.t. the MDL source structure $\widehat{S}$, which is an element in the class of tree sources of depth $D \leq \log (N/B)$, satisfies 
\begin{eqnarray*}
\rho_N(x^N) 
&\leq&  
B \left[ \log \left( \frac{N}{B}\right) + 2  \right]
\nonumber \\
&+&
B \left[ \frac{|\widehat{S_n}|}{2}\left[ \log \left(\frac{N}{B} \right) + 
O\left(1\right) \right] \right], 
\nonumber \\
\end{eqnarray*}
where $\widehat{S_n}$ is the estimated tree structure with the largest number of states among the $B$ tree structures.
\end{proposition}

\begin{proofoutline}
The redundancy analysis is similar to Theorem~\ref{theo1}, except that the model redundancy and parameter redundancy in the naive parallel compression algorithm are $B$ 
times the respective redundancies in the PTP-MDL algorithm as we estimate separate models for data of length $(N/B)$ in each of the $B$ PUs. $\blacksquare$

\end{proofoutline}

\subsection{Pass~I} \label{subsec:basic:phaseI}

{\bf Computing block symbol counts:}
Computational unit~$b$ computes $n^{\alpha}_s(b)$
for all $2^{D}$ depth-$D$ leaf contexts $s$.
In order for PU~$b$ to compute all block symbol counts in
$O(N/B)$ time, we define the {\em context index} $c_i(b)$
of the symbol $x_i(b)$ as
\begin{equation}
c_i(b) \triangleq
\sum_{j=0}^{D-1} 2^j x_{j+i-D}(b),
\label{eq:def:c_i(b)}
\end{equation}
where $i\in\{D+1,\ldots,N/B\}$ and
$x_{j+i-D}(b)\in\{0,1\}$,
hence $c_i(b)\in\{0,\ldots,2^{D}-1\}$.
Note that $c_i(b)$ is the binary number represented by
the context $s=x_{i-D}^{i-1}(b)$. Hence, it can
be used as a pointer to the address containing the
block symbol count $n_s^{\alpha}(b)$ for $s=x_{i-D}^{i-1}(b)$.
Moreover, the property
\begin{equation}
c_{i+1}(b)=
\frac{c_i(b)}{2}+
2^{D-1}x_i(b)-
\frac{x_{i-D}(b)}{2}
\label{eq:property:update_c_i}
\end{equation}
enables the computation of
all $N/B-D$ context indices of the symbols of $x(b)$ in $O(N/B)$
time complexity.

{\bf Constructing context trees:}
Because we restrict our attention to
depth-$D$ contexts, it suffices for PU~$b$ to compute
$\{n^{\alpha}_s(b)\}_{\alpha\in\cl{X},\ s\in\cl{X}^{D}}$,
all the block symbol counts of all the leaf contexts
of a full depth-$D$ context tree.
Information on internal nodes of the context
tree, whose depth is less than $D$, is computed
from the block symbol counts of the leaf contexts.

If $|s|=D$, then the CU gets
$\{n^{\alpha}_s(b)\}_{\alpha\in\cl{X}}$
from the PUs and computes $n_s^{\alpha}$
with (\ref{eq:n^alpha_s}).
Alternatively, $|s|<D$, and the CU
recursively derives $n_s^{\alpha}$
by adding up the symbol counts of children states, i.e.,
\begin{equation}
\label{eq:n^s_children}
n_s^{\alpha}= n_{0s}^{\alpha} + n_{1s}^{\alpha},
\qquad\forall\alpha\in\cl{X}.
\end{equation}

{\bf Computing the MDL source $\{\widehat{\cl{S}},\widehat{\Theta}\}$:}
\label{subsec:basic:phaseI:MDL}
For each state $s$, we either retain the children states
$0s$ and $1s$ in the tree or merge them into a single
state, according to which decision minimizes the coding length.
The coding length $l_s$ of the two-pass code that
describes the symbols generated by $s$ is
\begin{equation}
l_s =
\overbrace{\log(K_s)}^{\text{Pass~I}}
\overbrace{-n^0_s\log(1-r_s)-n^1_s\log(r_s)}^{\text{Pass~II}}.
\label{eq:l_s:binary}
\end{equation}

We now derive the coding length required for state $s$,
which is denoted by $\text{MDL}_s$.
For $|s|=D$, $n_s^0$ and $n_s^1$
are computed with (\ref{eq:n^alpha_s}),
$l_s$ is computed with (\ref{eq:l_s:binary}),
and $\text{MDL}_s=l_s$.
For $|s|<D$, we compute $n_s^\alpha$
hierarchically with (\ref{eq:n^s_children}),
after already having processed the children states.
In order to decide whether to prune the tree,
we compare $\text{MDL}_{0s}+\text{MDL}_{1s}$ with $l_s$.
Because retaining an internal node requires
the natural code~\cite{Willems1995CTW}
to describe that node (with $1$ bit),
\begin{equation*}
\text{MDL}_s=
\left\{
\begin{matrix}
l_s \quad \text{if $|s|=D$} \\
1+\min\left\{\text{MDL}_{0s}+\text{MDL}_{1s},
l_s \right\} & \text{else}
\end{matrix}
\right..
\end{equation*}
In terms of the natural code, if $|s|=D$, then
$s$ is a leaf of the full depth-$D$ context tree,
and its natural code is empty; else $|s|<D$,
and the natural code requires $1$ bit to encode whether
$s\in\cl{S}$.
The symbols generated by $s$ are  encoded either
by retaining the children states (this requires
a coding length of $\text{MDL}_{0s}+\text{MDL}_{1s}$ bits),
or by pruning the children states and retaining
state $s$ with coding length $l_s$.
If $|s|=D$, then we do not process deeper contexts.
The context tree pruning has $O(N/B)$ time complexity because the tree has $O(N/B)$ states.

\subsection{Pass~II}

In Pass~II, PU~$b$ knows $\widehat{\cl{S}}$ and
$\{r_s\}_{s\in\widehat{\cl{S}}}$.
PU~$b$ encodes $x(b)$ sequentially;
for each symbol $x_i(b)$, it determines 
the generator state $G_i(b)$.
An $O(N/B)$ algorithm  for determining
$G_i(b)$ for all the symbols of $x(b)$ utilizing (\ref{eq:def:c_i(b)},\ref{eq:property:update_c_i}) is described by Baron~\cite{BaronThesis}.
After determining $G_i(b)$, the symbol $x_i(b)$
is encoded according to the probability assignment
$\widehat{p}(x_i(b))\triangleq\widehat{\theta}_{G_i(b)}^{x_i(b)}$
with an arithmetic encoder~\cite{Cover06}.
In order to have $O(N/B)$ time complexity and $O(1)$ expected coding redundancy per PU,
arithmetic coding is performed with $\log(N)$ bits of
precision~\cite{Cover06}, where we assume that the hardware architecture performs arithmetic with $\log(N)$ bits of precision in $O(1)$ time.

\subsection{Decoder} \label{subsec:basic:decoder}

The $B$ decoding blocks can be implemented on $B$ PUs.
Decoding block~$b$ decodes $x(b)$ sequentially;
for each symbol $x_i(b)$, it determines $G_i(b)$.
The same $O(N/B)$ algorithm used in Pass~II for
determining $G_i(b)$ for all the symbols of
$x(b)$ can be used in the $B$ decoding blocks.
After determining $G_i(b)$, the symbol $x_i(b)$
is decoded according to the probability assignment
$\widehat{p}(x_i(b))\triangleq\widehat{\theta}_{G_i(b)}^{x_i(b)}$
with an arithmetic decoder~\cite{Cover06} 
that has $O(N/B)$ time complexity.
We have the following result for the overall time
complexity of the encoder and decoder.

\begin{theorem}\label{theo2}(Theorem 16 in Baron~\cite{BaronThesis}.)
With computations performed with $\log(N)$ bits of precision defined as $O(1)$ time, the PTP-MDL encoder and decoder each require $O(N/B)$ time.
\end{theorem}

\begin{proofoutline}
(For a detailed presentation, see pages 116 -- 120 in Baron~\cite{BaronThesis}.) For the algorithm to run in $O(N/B)$ time, all subroutines in Pass~I and Pass~{II} should run in $O(N/B)$ time. An outline of the time complexity analysis for the subroutines is as follows.

\begin{itemize}
\item Computing the block symbol counts in (\ref{eq:def:n_s_alpha_b}) has $O(N/B)$ complexity as $x^N$ is divided among $B$ PUs, and the context index of each symbol in $x^N$ can be updated in $O(1)$ time utilizing (\ref{eq:def:c_i(b)},\ref{eq:property:update_c_i}).
\item Adding up the block symbol counts in (\ref{eq:n^alpha_s}) for each state can be performed in O(1) time using a pipelined adder tree~\cite{BaronThesis}. Since we limit the maximum number of states to be $2^D = (N/B)$, the computational complexity is $O(N/B)$.
\item The context tree pruning processes $O(2^D) = O(N/B)$ contexts, and each function call decides to prune based on O(1) computations involving equations (\ref{eq:n^s_children}, \ref{eq:l_s:binary}).
\item A generator look up table, which maps the generator state $G_i(b)$ to the probability assignment $\widehat{p}(x_i(b))$, is constructed in $O(N/B)$ time utilizing  (\ref{eq:def:c_i(b)},\ref{eq:property:update_c_i}).
\item Pass~II has $O(N/B)$ time complexity when context indices and a generator look up table are used to assign probabilities to each of the $B$ arithmetic encoders. The decoder also has $O(N/B)$ time complexity, because its structure is similar to that of Pass~II.
\end{itemize}
In combination, these steps yield the $O(N/B)$ time complexity result of Theorem~\ref{theo2}. $\blacksquare$
\end{proofoutline}

\subsection{Quantization schemes for better compression} \label{subsec:basic:improv}

In equation~(\ref{eq:K^*}) for estimating $K_s$, we used a conservative estimate for each {\em context population}, 
which is defined as the number of symbols for a given context, 
i.e., $n_s = n_s^0 + n_s^1$, 
where we implicitly assumed $n_s = N$.
However, the number of symbols for a particular 
context is often much smaller than $N$. 
We can improve the compression by encoding the bin index of each
context
using the corresponding context population in (\ref{eq:K^*}). 
For the above scheme, we may need to spend extra bits to represent 
the size of each context population.

We propose two quantization schemes for better compression in which we first prune the context tree using (\ref{eq:K^*}). In our first scheme, we use
roughly $1.772 (N/|\widehat{\cl{S}}|)^{0.5}$
 quantization bins for encoding each bin index. With this scheme, there is no need to encode context populations,
because the quantizer will perform well on average. 
In our second scheme, we use a 2-level quantization 
scheme in which all contexts whose population is below 
some threshold $\tau$ are processed with a small coarse 
quantizer whose index can be encoded with a minimal 
number of bits, while contexts whose population 
exceeds $\tau$ are processed with a larger fine quantizer.
Both of our quantization schemes have only a minor impact on the
run time over the original quantization scheme (\ref{eq:K^*}). 
Note that Theorems 1 and 2 hold for the modifications we propose for the two new quantizers.

Using a reduced-length quantizer might encourage the MDL
optimization to increase the number of states, which could
further improve the compression. Indeed, we tried to rerun
the MDL optimization procedure with
$N/|\widehat{\cl{S}}|$ instead of $N$ in (\ref{eq:K^*}), 
but this increased the run time by 
$30-40\%$ while yielding modest compression gains.
Therefore, we do not recommend rerunning the MDL optimization. 

Vector quantization~\cite{GershoGray1993,Cover06}, which can be used when the binary symbols are clustered to form a larger alphabet, has the potential to improve the compression performance by up to an additive constant of $O\left(\frac{K}{2}\log(K)\right)$~\cite{XieBarron1997,Beirami2011} over the scalar quantization based compression algorithm. However, in order to maintain $O(N/B)$ time complexity, we may need to reduce the maximum depth $D$.

\section{Numerical Results} \label{sec:num_res}

\newcommand {\specialcell} [2][c]{
\begin{tabular} [#1] {@{}c@{}}#2 \end{tabular}}

\newcommand {\threecell} [3][c]{
\begin{tabular} [#1] {@{}c@{}c@{}}#2 \end{tabular}}

\renewcommand{\tablename}{Table}

\begin{table} [t]
\vspace*{+2.5mm}
\centering
\begin{tabular} {|c|c| c| c| c|}
\hline 
\multirow{2}{*}{\specialcell{Compression\\ algorithm}} & \multicolumn{3}{c|}{\specialcell{Compression ratio\\ (bits/byte)}} & \multirow{2}{*}{\threecell{Average\\ throughput\\ (Mbps)}}\\
\hhline{~---}
 & E.Coli & bible.txt &\specialcell{ world \\192.txt} & \\
\hline
LZ77a (32KB) & 2.35 & 2.32 & 2.32 &31 \\
LZ77b (4MB) & 2.27 & 1.93 & 1.72 &36 \\
BWT & 2.16 & 1.67 & 1.58 &36 \\
NanoZip &1.97 & 1.42&1.25 &84 \\
Gipfeli & 6.00&7.49 &7.27 & 826\\
LZ4 & 5.45 & 4.14 & 3.98 &1515\\
Snappy & 3.73&3.93 &4.02 &2025 \\
\hline
PTP-MDL (B=1) & 1.98 & 2.15 & 2.45 &2\\
-"- \ \ (B=10) & 1.99 & 2.36 & 2.85 &17\\
-"- \ (B=100) & 1.99 & 2.57 & 3.20 &138\\
-"- (B=1000) & 2.01 & 3.09 & 3.77 &908\\
\hline
\end{tabular}
\caption{Performance comparison for different compressors.} 
\label{comp_perf}
\end{table}

Having described the PTP-MDL algorithm, we have set the stage to describe our 
numerical results with a prototype implementation~\cite{KrisBaronSIP2014}. 
After surveying our simulation setting, 
we compare the compression ratio and the throughput of the PTP-MDL algorithm 
with two types of algorithms: ({\em i}) high quality compressors
and ({\em ii}) fast 
compression algorithms. The PTP-MDL algorithm offers competitive performance with
both types of algorithms. We show the trade-off between compression ratio and 
throughput of the PTP-MDL algorithm as a function of $B$. 
Finally, we show the improvement in compression performance of the PTP-MDL
algorithm using the quantization schemes proposed 
in Section~\ref{subsec:basic:improv}.

\subsection{Simulation setting} 

\begin{figure}[t]
\vspace*{-5mm}
\includegraphics[angle=0,width=8cm]{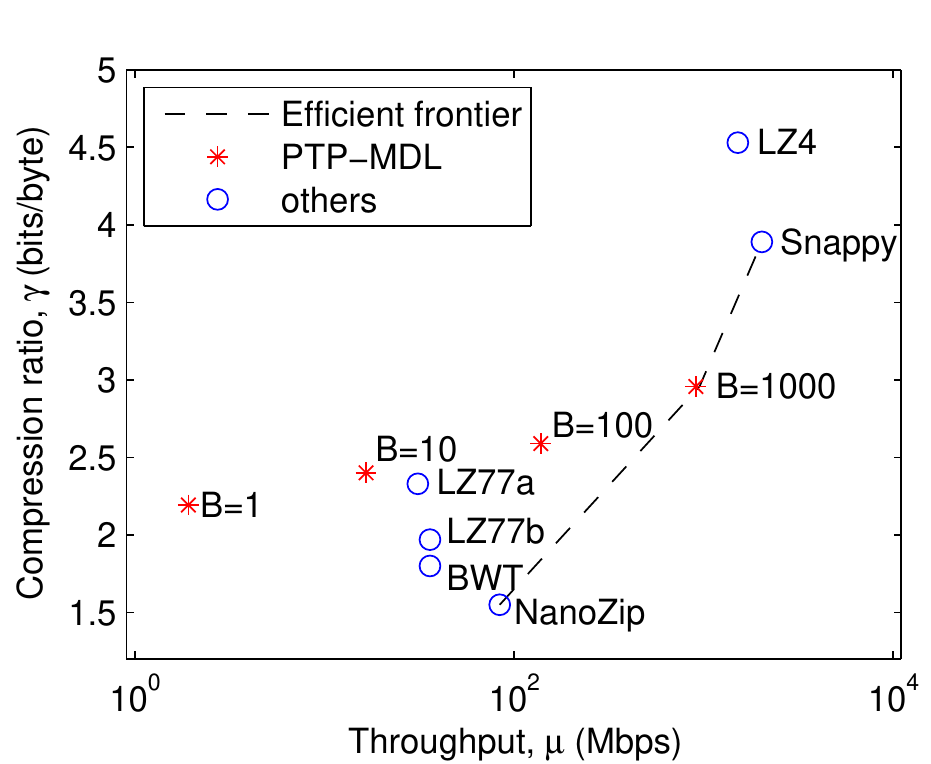}
\caption
{Compression ratio vs. throughput for different compressors.}
\label{fig:CR_data_rate}
\end{figure}

We have developed a serial (non-parallel) implementation in C++, which serves as a prototype that allows us to evaluate anticipated performance of a GPU implementation, which is ongoing work. The algorithm treats any data as a binary bit stream; in future work, we plan to apply techniques that exploit the byte nature of real-world data~\cite{Willems1998}. Although the parallel parts of the algorithm run sequentially in the current implementation, we give a predicted time performance as

\begin{equation}
t_{\text{est}}(B) = t_{\text{s}} +\frac{ t_{\text{sp}}}{\eta B},
\label{eqn:Amdahl}
\end{equation}
where $t_{\text{est}}(B)$ is the estimated time for executing the algorithm using $B$ PUs, $t_{\text{s}}$ is the time required to execute the serial part of the algorithm, $ t_{\text{sp}}$ is the sequential time required for executing the parallel part of the algorithm, and $\eta \in [0,1]$ is the {\em efficiency of parallelization}. 

The {\em compression ratio}, $\gamma$, is defined as the average number of output bits in the compressed data required to represent one input byte of uncompressed data. Note that the lower the $\gamma$, the better the compression. The {\em throughput}, $\mu = \frac{N}{t_{\text{est}}(B)}$, is measured in megabits per second (Mbps); the higher the throughput, the shorter the run time. We assume $\eta = 0.2$ in our simulations to provide a conservative estimate of the throughput unless specified otherwise.

\subsection{PTP-MDL vs. other algorithms} 

\begin{figure}[t]
\vspace*{-2.5mm}
\includegraphics[angle=0,width=8cm]{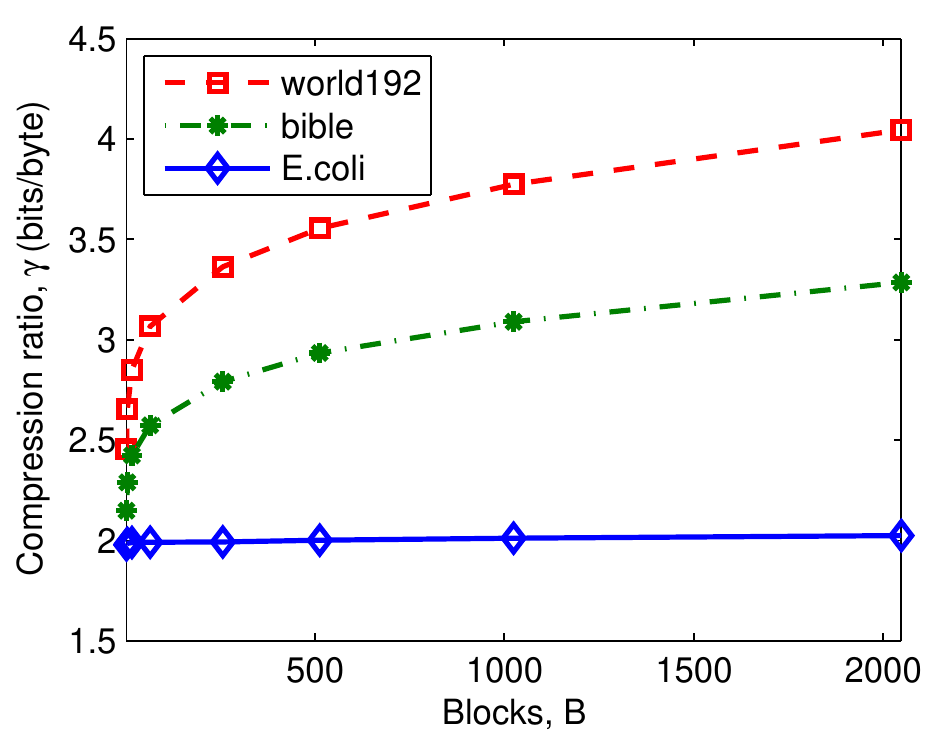}
\caption
{Compression ratio vs. number of blocks.}
\label{fig:CR_B}
\end{figure}

We compare the compression ratio and throughput of several 
algorithms for files from the large Canterbury corpus 
(http://corpus.canterbury.ac.nz/descriptions) in 
Table~\ref{comp_perf} and Figure~\ref{fig:CR_data_rate}. 
The best trade-off is achieved for algorithms that have the lowest compression ratio and the highest throughput, and are represented by points that are closest to the bottom right corner in Figure~\ref{fig:CR_data_rate}.
We ran the PTP-MDL algorithm 
for $B \in \{1, 10, 100, 1000\}$. Although 
running $B=1000$ parallel units may seem ambitious, 
GPUs with thousands of cores are already commonplace.
PTP-MDL is compared with high quality compressors 
such as Lempel-Ziv coding~\cite{Cover06} with a 32KB dictionary size (LZ77a), Lempel-Ziv coding with a 4MB dictionary size (LZ77b), the Burrows-Wheeler transform (BWT)~\cite{BurrowsWheeler}, 
and the context-based NanoZip algorithm 
(http://nanozip.net/), 
as well as fast algorithms such as LZ4~\cite{LZ4}, Gipfeli~\cite{Lenhardt2012}, and Snappy~\cite{Snappy}. For the E.coli file, 
PTP-MDL and NanoZip have the lowest compression ratio. Compression with NanoZip also provides the lowest compression ratios for the two text files, 
bible.txt and
world192.txt. 

The PTP-MDL algorithm has a competitive compression ratio
although this algorithm is implemented for binary symbols, whereas the other
algorithms are designed for 8 bit symbols,
to the best of our knowledge. LZ4, Gipfeli, and Snappy, which are speed optimized approximations of LZ77~\cite{Cover06}, did not do well in compression ratio performance. 
In addition to a competitive compression ratio, as $B$ increases,
the throughput of the PTP-MDL algorithm is comparable to the throughputs of fast data compression algorithms such as 
LZ4 ($\approx1515$ Mbps) and Snappy ($\approx2025$ Mbps) (Table~\ref{comp_perf}), which can be used in big data problems. 

We highlight that the efficient frontier in 
Figure~\ref{fig:CR_data_rate} reflects the optimal trade-off between 
compression ratio and throughput. 
PTP-MDL extends the efficient frontier for large $B$.

\subsection{Compression ratio and throughput vs. $B$} 

{\bf PTP-MDL ${\bf \gamma}$ performance vs. B:} Figure~\ref{fig:CR_B} illustrates the impact of the number of blocks $B$ on the compression ratio $\gamma$. For a simple source such as E.coli, where $D$ is small, the compression ratio increase with $B$ is modest. However, there is a non-linear behavior for more complicated data such as English text. For bible.txt and world192.txt, the compression ratio $\gamma$ deteriorates rapidly for small $B$. This deterioration could be due to the decrease in maximum depth available for the tree source given by $O(\log(N/B))$, which may impact the coding length as $B$ increases. 

\begin{figure}[t]
\vspace*{-2.5mm}
\includegraphics[angle=0,width=8cm]{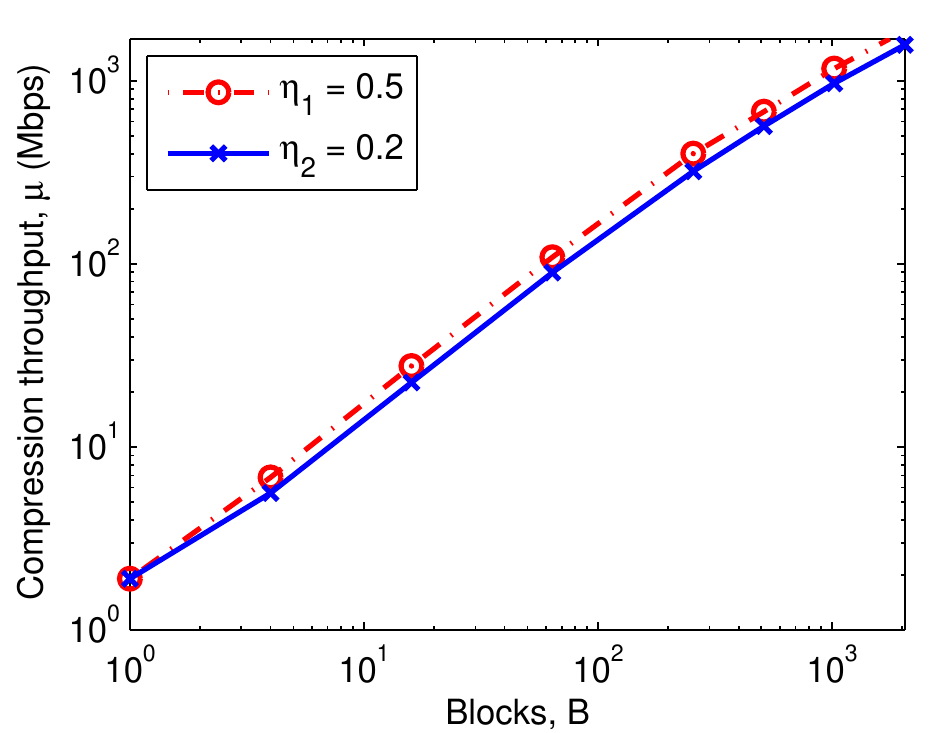}
\caption
{Compression throughput vs. number of blocks.}
\label{fig:comp_data_rate}
\end{figure}

{\bf PTP-MDL ${\bf \mu}$ performance vs. B:} Figure~\ref{fig:comp_data_rate} illustrates the impact of the number of blocks $B$ on the average compression  throughput $\mu$ for files from the large Canterbury corpus for two efficiencies of parallelization, $\eta_1 = 0.5$ and $\eta_2 = 0.2$.  For small $B$, the speed up is linear, and as $B$ increases, the speed up slows down. This trend is due to {\em Amdahl's law}~\cite{Solihin2009} given in (\ref{eqn:Amdahl}). When $B$ is low, the fraction of serial execution time is insignificant compared to parallel execution time. However, when $B$ is large, the parallel fraction of execution time is reduced, which introduces nonlinearity in the throughput $\mu$ for large values of $B$. The result for decompression throughput is similar to that of compression throughput, and thus omitted for brevity. 

\subsection{Compression ratio for 2 level quantization scheme} 

Table~\ref{2quant_comp_perf} shows percentage improvements 
in the compression ratio $\gamma$
obtained using the two quantization schemes discussed in 
Section~\ref{subsec:basic:improv}
w.r.t. the original quantization scheme (\ref{eq:K^*})
for files from the large Canterbury corpus. 
Among the two quantization schemes, the second scheme based on 
2-level quantization narrowly outperforms the first scheme.
From Table~\ref{2quant_comp_perf}, it can be seen that 
the E.coli file, which PTP-MDL already compresses almost
as well as the state of the art NanoZip (Table~\ref{comp_perf}), 
does not improve much. 
For the text files bible.txt and world192.txt, 
the 2-level quantization scheme typically improves compression
by $1-2\%$. Note, however, that the compression improvement is 
modest for large $B$.

\section{Conclusion}\label{sec:conclusion}

In this paper, we have proposed a parallel two-pass minimum 
description length (PTP-MDL) algorithm for compressing 
context tree sources of depth $D = \log (N/B)$. The algorithm can 
compress and decompress in $O(N/B)$ time while achieving a redundancy within 
$B\log (N/ B)$ bits of Rissanen's lower bound on 
universal compression performance~\cite{Rissanen1996}. 
We further improved the PTP-MDL algorithm using a 2 level quantization scheme. In future work, we plan to explore and analyze the use of multi-level quantizers to improve the compression performance without degrading the time complexity.

Our numerical results show that the compression ratio $\gamma$ of our algorithm for real-world data is comparable to existing universal data compressors. Moreover, the throughput of PTP-MDL scales well with the number of parallel units, even for large $B$. We speculate that exploiting the byte nature of 
real-world data can further improve the compression ratio $\gamma$.

The PTP-MDL algorithm has the useful property of random access~\cite{Kreft2010}, where any part of the compressed file can be decompressed without decompressing the entire file. This property is useful in applications where the compressed file is large, and only part of the file is needed to service a query. The JPEG2000 compression standard uses the random access property of the Embedded Block Coding with Optimized Truncation (EBCOT) based lossless encoder to selectively reconstruct the region of interest~\cite{Taubman2000}. The PTP-MDL algorithm can replace the lossless encoder in JPEG2000 to improve the throughput without losing the random access property. It might be possible to also apply PTP-MDL to lossless image compression using the two-dimensional contexts of Weinberger et al.~\cite{Weinberger1996}, but in the image setting the number of contexts is often prohibitively large, and it is not clear whether comprehensive changes to PTP-MDL would be needed.

Finally, although the two-pass approach may seem costly in applications where data is streamed (online compression), we can design the compression system that minimizes the impact of redundancy by dividing the data into reasonably large blocks such that the entire block can be processed simultaneously in random access memory (RAM). 
With currently available hardware circa 2015, 
our proposed algorithm has the potential to compress data of the order of gigabytes (GBs) simultaneously. 
Another related direction for future inquiry is to lower memory utilization in the parallel compression design space~\cite{Beirami2012}.
 
\begin{table} [t]
\vspace*{+2.5mm}
\centering
\begin{tabular} {|c|c|c|c|c|c|c| }
\hline 
\multirow{2}{*}{B}& \multicolumn{3}{c|}{Quantization scheme 1}& \multicolumn{3}{c|}{Quantization scheme 2}\\
\hhline{~------}
&\specialcell{E \\.Coli}&\specialcell{bible \\.txt} &\specialcell{ world \\192.txt} &\specialcell{E \\.Coli}&\specialcell{bible \\.txt} &\specialcell{ world \\192.txt}\\
\hline
1 & 0.03\% & 1.79\% & 4.37\%& 0.01\% & 1.91\% & 4.72\% \\
10 & 0.01\% & 1.09\% & 2.28\%& 0.01\% & 1.14\% & 2.50\% \\
100 & 0.01\% & 0.58\% & 1.05\%& 0.01\% & 0.64\% & 1.35\% \\
1000& 0.00\% & 0.16\% & 0.37\% & 0.00\% & 0.21\% & 0.44\% \\
\hline
\end{tabular}
\caption{Percentage improvements in the compression ratio $\gamma$ obtained using quantization schemes in Section~\ref{subsec:basic:improv} w.r.t. the original quantization scheme (\ref{eq:K^*}).} 
\label{2quant_comp_perf}
\end{table}

\section*{Acknowledgements}
We thank Yoram Bresler and Mehmet K{\i}van\c{c} M{\i}h\c{c}ak for numerous discussions relating to this work;  Frans Willems for the arithmetic code implementation;
and the reviewers, the editor, Yanting Ma, Jin Tan, and Junan Zhu for their careful evaluation of the manuscript.

\bibliographystyle{IEEEbib}
\bibliography{cites}

\begin{thebibliography}{10}

\bibitem{KrisBaron2014}
N.~Krishnan, D.~Baron, and M.~K. M{\i}h\c{c}ak,
\newblock ``A parallel two-pass {MDL} context tree algorithm for universal
  source coding,''
\newblock in {\em Proc. Int. Symp. Inf. Theory (ISIT)}, July 2014.

\bibitem{KrisBaronSIP2014}
N.~Krishnan and D.~Baron,
\newblock ``Performance of parallel two-pass {MDL} context tree algorithm,''
\newblock in {\em Proc. IEEE Global Conf. Signal Inf. Process.}, Atlanta, GA,
  Dec. 2014.

\bibitem{ArmingIBWT}
S.~Arming, R.~Fenkhuber, and T.~Handl,
\newblock ``Data compression in hardware -- the {B}urrows-{W}heeler approach,''
\newblock in {\em IEEE Int. Symp. Des. Diagnostics Electron. Circuits Syst.},
  Apr. 2010, pp. 60--65.

\bibitem{Lenhardt2012}
R.~Lenhardt and J.~Alakuijala,
\newblock ``Gipfeli - high speed compression algorithm,''
\newblock in {\em Proc. Data Compression Conference (DCC)}, Apr. 2012, pp.
  109--118.

\bibitem{Snappy}
S.~Gunderson,
\newblock ``{Snappy, A fast compressor/decompressor},''
\newblock code.google.com/p/snappy/.

\bibitem{QuickLZ}
L.~M. Reinhold,
\newblock ``{QuickLZ website},''
\newblock http://www.quicklz.com/.

\bibitem{FastLZ}
A.~Hidayat,
\newblock ``{FastLZ website},''
\newblock http://fastlz.org/.

\bibitem{FRT96}
P.~Franaszek, J.~Robinson, and J.~Thomas,
\newblock ``Parallel compression with cooperative dictionary construction,''
\newblock in {\em Proc. Data Compression Conf. (DCC)}, Mar. 1996.

\bibitem{Beirami2012}
A.~Beirami and F.~Fekri,
\newblock ``On lossless universal compression of distributed identical
  sources,''
\newblock in {\em Proc. Int. Symp. Inf. Theory (ISIT)}, July 2012, pp.
  561--565.

\bibitem{ST2001}
M.~L.~A. Stassen and T.~J. Tjalkens,
\newblock ``A parallel implementation of the {CTW} compression algorithm,''
\newblock in {\em Proc. 22d Benelux Symp. Inf. Comm.}, May 2001, pp. 85--92.

\bibitem{Willems1995CTW}
F.~M.~J. Willems, Y.~M. Shtarkov, and T.~J. Tjalkens,
\newblock ``The context tree weighting method: {B}asic properties,''
\newblock {\em IEEE Trans. Inf. Theory}, vol. 41, no. 3, pp. 653--664, May
  1995.

\bibitem{Cover06}
T.~M. Cover and J.~A. Thomas,
\newblock {\em Elements of Information Theory},
\newblock New York, NY, USA: Wiley-Interscience, 2006.

\bibitem{Willems2000}
F.~M.~J. Willems,
\newblock ``Some challenges in source coding,''
\newblock in {\em Proc. 3rd ITG Conf. Source Channel Coding}, Jan. 2000, pp.
  245--249.

\bibitem{Rissanen1978}
J.~Rissanen,
\newblock ``Modeling by shortest data description,''
\newblock {\em Automatica}, vol. 14, no. 5, pp. 465--471, Sept. 1978.

\bibitem{CLR}
T.~H. Cormen, C.~E. Leiserson, and R.~L. Rivest,
\newblock {\em Introduction to Algorithms},
\newblock The MIT Press, Cambridge, MA, 2009.

\bibitem{Kreft2010}
S.~Kreft and G.~Navarro,
\newblock ``{LZ}77-like compression with fast random access,''
\newblock {\em Data Compression Conference (DCC), 2010}, pp. 239--248, Mar.
  2010.

\bibitem{BaronThesis}
D.~Baron,
\newblock ``{F}ast parallel algorithms for universal lossless source coding,''
\newblock Feb. 2003,
\newblock {P}h.D. thesis, UIUC.

\bibitem{WMF1994}
M.~J. Weinberger, N.~Merhav, and M.~Feder,
\newblock ``{Optimal sequential probability assignment for individual
  sequences},''
\newblock {\em IEEE Trans. Inf. Theory}, vol. 40, no. 2, pp. 384--396, Mar.
  1994.

\bibitem{Rissanen1996}
J.~Rissanen,
\newblock ``{Fisher information and stochastic complexity},''
\newblock {\em IEEE Trans. Inf. Theory}, vol. 42, no. 1, pp. 40--47, Jan. 1996.

\bibitem{BWTMDL}
D.~Baron and Y.~Bresler,
\newblock ``An {O(N)} semipredictive universal encoder via the {BWT},''
\newblock {\em IEEE Trans. Inf. Theory}, vol. 50, no. 5, pp. 928--937, May
  2004.

\bibitem{Volf1995CTWMDL}
P.~A.~J. Volf and F.~M.~J. Willems,
\newblock ``A study of the context tree maximizing method,''
\newblock in {\em Proc. 16th Benelux Symp. Inf. Theory, Nieuwerkerk Ijsel,
  Netherlands}, May 1995, pp. 3--9.

\bibitem{WillemsCISS}
F.~M.~J. Willems, Y.~M. Shtarkov, and T.~J. Tjalkens,
\newblock ``Context-tree maximizing,''
\newblock in {\em Proc. Conf. Inf. Sci. Syst.}, Mar. 2000, pp. 7--12.

\bibitem{BaronBreslerMihcak2003}
D.~Baron, Y.~Bresler., and M.~K. M{\i}h{\c{c}}ak,
\newblock ``Two-part codes with low worst-case redundancies for distributed
  compression of {B}ernoulli sequences,''
\newblock in {\em Proc. Conf. Inf. Sciences Systems}, Mar. 2003.

\bibitem{GershoGray1993}
A.~Gersho and R.~M. Gray,
\newblock {\em {Vector quantization and signal compression}},
\newblock Kluwer, 1993.

\bibitem{XieBarron1997}
Q.~Xie and A.~R. Barron,
\newblock ``Minimax redundancy for the class of memoryless sources,''
\newblock {\em IEEE Trans. Inf. Theory}, vol. 43, no. 2, pp. 646--657, Mar.
  1997.

\bibitem{Beirami2011}
A.~Beirami and F.~Fekri,
\newblock ``Results on the redundancy of universal compression for
  finite-length sequences,''
\newblock in {\em Proc. Int. Symp. Inf. Theory (ISIT)}, July 2011, pp.
  1504--1508.

\bibitem{Willems1998}
F.M.J. Willems,
\newblock ``{The context-tree weighting method: Extensions},''
\newblock {\em IEEE Trans. Inf. Theory}, vol. 44, no. 2, pp. 792--798, Mar.
  1998.

\bibitem{BurrowsWheeler}
M.~Burrows and D.J. Wheeler,
\newblock {\em {A block-sorting lossless data compression algorithm}},
\newblock 1994.

\bibitem{LZ4}
``{LZ4, Extremely Fast Compression algorithm},''
\newblock code.google.com/p/lz4/.

\bibitem{Solihin2009}
Y.~Solihin,
\newblock {\em Fundamentals of Parallel Computer Architecture},
\newblock Solihin Publishing and Consulting LLC, 2009.

\bibitem{Taubman2000}
D.~Taubman,
\newblock ``High performance scalable image compression with {EBCOT},''
\newblock {\em IEEE Trans. Image Process.}, vol. 9, no. 7, pp. 1158--1170, Jul.
  2000.

\bibitem{Weinberger1996}
M.J. Weinberger, J.J. Rissanen, and R.B. Arps,
\newblock ``Applications of universal context modeling to lossless compression
  of gray-scale images,''
\newblock {\em IEEE Trans. Image Process.}, vol. 5, no. 4, pp. 575--586, Apr.
  1996.

\end{thebibliography}

\begin{IEEEbiography}[{\includegraphics[width=1in,height=1.25in,clip,keepaspectratio]{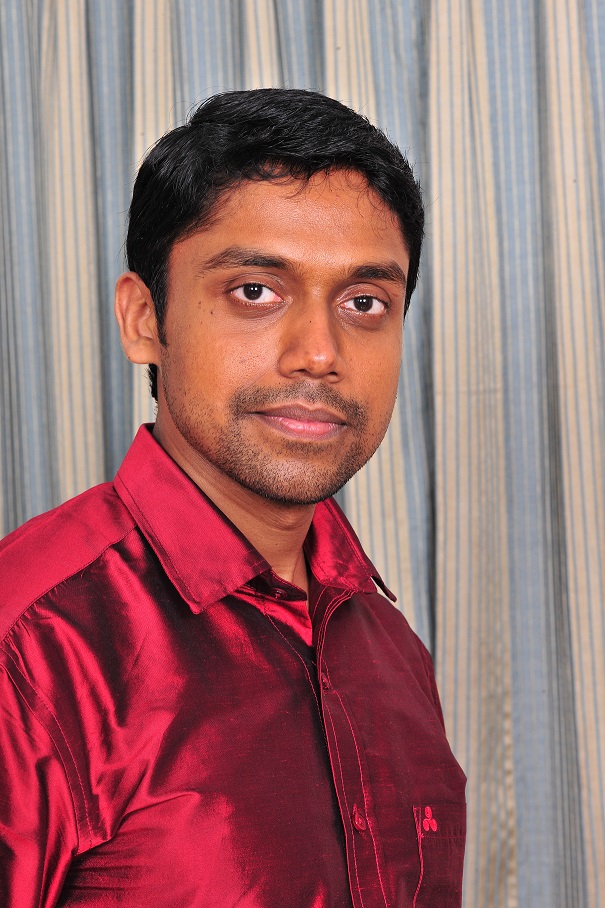}}]
{Nikhil Krishnan} (S'13) was born in Trivandrum, India, in 1984. He received the B.Tech. degree in electronics and communication engineering from the College of Engineering Trivandrum, India, in 2006, and the M.Tech. degree in electrical engineering from the Indian Institute of Technology (IIT) Bombay, India, in 2010. 

From 2006 to 2008, he worked as a Scientist at the Indian Space Research Organization, India, in navigation systems. From 2010 to 2012, he worked at Samsung India Software Operations 
in image processing. Since 2012, he has been pursuing his Ph.D. in the 
Electrical and Computer Engineering Department at NC State University in Raleigh. His current research interests include information theory, signal processing, compression, and performance of algorithms.  
\end{IEEEbiography}

\begin{IEEEbiography}[{\includegraphics[width=1in,height=1.25in,clip,keepaspectratio]{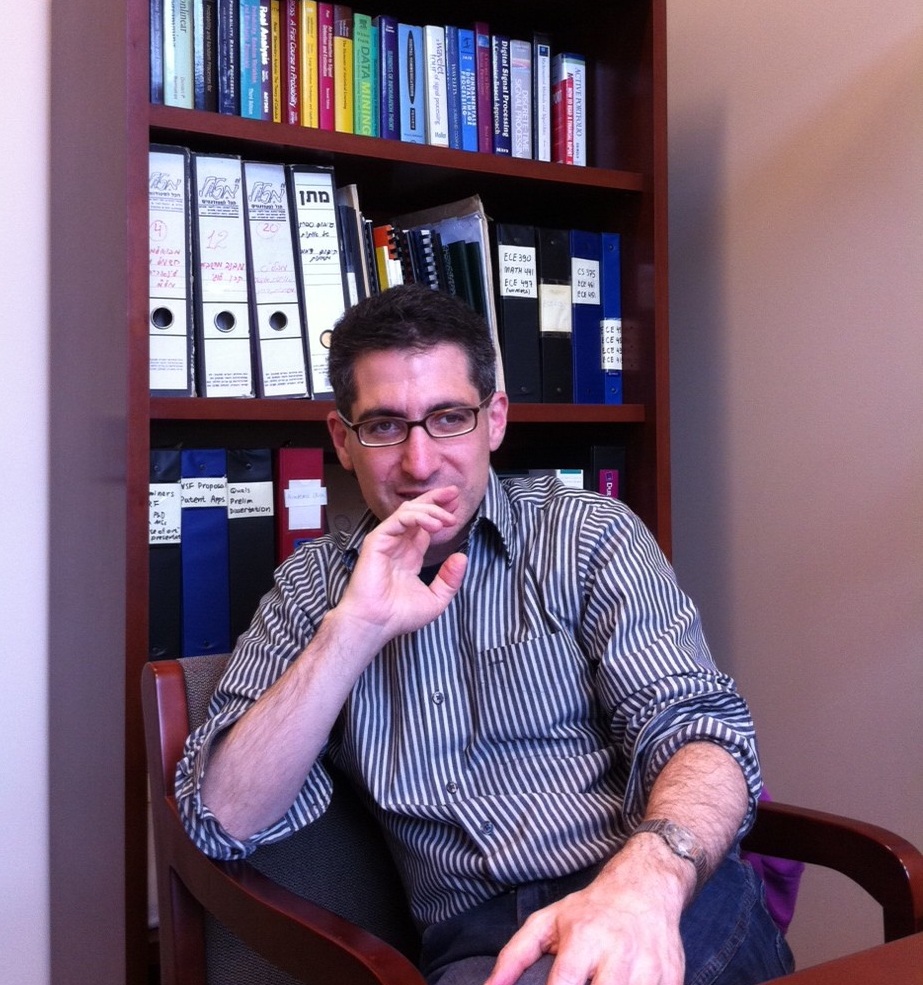}}]
{Dror Baron} (S'99-M'03-SM'10) received the B.Sc. (summa cum laude) and
M.Sc. degrees from the Technion - Israel Institute of Technology, Haifa, Israel,
in 1997 and 1999, and the Ph.D. degree from the University of Illinois at Urbana-Champaign 
in 2003, all in electrical engineering.
From 1997 to 1999, Dr. Baron worked at Witcom Ltd. in modem design.
From 1999 to 2003, he was a research assistant at the University of Illinois at
Urbana-Champaign, where he was also a Visiting Assistant Professor in 2003.
From 2003 to 2006, he was a Postdoctoral Research Associate in the Department
of Electrical and Computer Engineering at Rice University, Houston, TX. From
2007 to 2008, he was a quantitative financial analyst with Menta Capital, San
Francisco, CA, and from 2008 to 2010 he was a Visiting Scientist in the Department 
of Electrical Engineering at the Technion - Israel Institute of Technology,
Haifa. Since 2010, Dr. Baron has been an Assistant Professor in the Electrical
and Computer Engineering Department at NC State University.
Dr. Baron's research interests combine information theory, signal processing,
and fast algorithms; in recent years, he has focused on compressed sensing. 

Dr. Baron was a recipient of the 2002 M. E. Van Valkenburg Graduate Research
Award, and received honorable mention at the Robert Bohrer Memorial Student
Workshop in April 2002, both at the University of Illinois. He also participated
from 1994 to 1997 in the Program for Outstanding Students, comprising the top
0.5\% of undergraduates at the Technion.
\end{IEEEbiography}
\end{document}